\documentclass[journal=jpccck,manuscript=letter]{achemso}

\usepackage{graphicx}
\usepackage{graphics}
\usepackage{amsmath}
\usepackage{amsfonts}
\usepackage{color}
\usepackage{caption}
\usepackage{subcaption}
\usepackage{siunitx}

\title{Computational Amperometry of Nanoscale Capacitors in Molecular Simulations}  
\author{Thomas Dufils}
\affiliation[SU]{Sorbonne Universit\'e, CNRS, Physico-chimie des Electrolytes et Nanosyst\`emes Interfaciaux, PHENIX, F-75005 Paris, France}
\author{Michiel Sprik}
\affiliation[Cambridge]{Yusuf Hamied Department of Chemistry, University of Cambridge, Cambridge CB2 1EW, United Kingdom}
\author{Mathieu Salanne}
\affiliation[SU]{Sorbonne Universit\'e, CNRS, Physico-chimie des Electrolytes et Nanosyst\`emes Interfaciaux, PHENIX, F-75005 Paris, France}
\alsoaffiliation[IUF]{Institut Universitaire de France (IUF), 75231 Paris Cedex 05, France}
\alsoaffiliation[RS2E]{R\'eseau sur le Stockage Electrochimique de l'Energie (RS2E), FR CNRS 3459, France}
\email{mathieu.salanne@sorbonne-universite.fr}
\date{\today}

\begin{document}

\begin{abstract}
	In recent years, constant applied potential molecular dynamics has allowed to study the structure and dynamics of the electrochemical double-layer of a large variety of nanoscale capacitors. Nevertheless it remained impossible to simulate polarized electrodes {\it at fixed total charge}. Here we show that combining a constant potential electrode with a finite electric displacement fills this gap by allowing to simulate open circuit conditions. The method can be extended by applying an electric displacement ramp to perform computational amperometry experiments at different current intensities. As in experiments, the full capacitance of the system is obtained at low intensity, but this quantity decreases when the applied ramp becomes too fast with respect to the microscopic dynamics of the liquid.  
\end{abstract}

\maketitle


Recent methodological advances have allowed the study of electrified interfaces by molecular simulations. On the one hand, electrochemical systems made of an electrolyte surrounded by two metallic electrodes are now routinely simulated using the constant applied potential method, in which a potential constraint is applied on a set of auxiliary dynamic variables (the electrode atom partial charges are most often used for the latter)~\cite{siepmann1995a}. In order to comply with the peculiar geometry of the system, it is possible to use 2-dimensional periodic boundary conditions (PBCs) using a modified Ewald treatment of electrostatic interactions~\cite{reed2007a}. This approach was mostly used in the absence of redox reactions, {\it i.e.} to study the structure, the dynamics and the capacitance of the electrochemical double-layer formed in nanoscale capacitors~\cite{merlet2013b,vatamanu2013a,burt2016a,bi2020a}, albeit recent developments allowed to study the thermodynamics of electron transfer reactions in the vicinity of electrodes~\cite{li2017j,dwelle2019a,limaye2020a,bhullar2021a}. On the other, the finite field extended Lagrangian approach~\cite{zhang2016f}
has been  used in several classical and {\it ab initio} molecular dynamics (MD) studies of aqueous interfaces~\cite{sayer2019a,zhang2019a} (for a recent review see Ref.~\citenum{zhang2020a}). These simulations are carried in a fully periodic 3-dimensional MD cell using standard Ewald summation methods. The finite field MD method, originally based on thermodynamic arguments~\cite{stengel2007a}, was subsequently reformulated in a dynamical Lagrangian framework~\cite{sprik2018a} and applied in  a study transport properties in a model ionic aqueous solution~\cite{cox2019a}. 

Recently, we have merged the constant potential and finite field method in a design of  a periodic single electrode electrochemical cell~\cite{dufils2019a}. The electrode is polarized by the field, yielding two electrochemical interfaces with opposite charges on its two sides. The electrolyte responds with compensating layers of excess charge leading to the formation of two electric double layers in series. The potential steps across these double layers add up to a net potential across the MD cell.  Electrode slab and electrolyte each remain overall neutral for a non-Faradaic interface (no electrode -electrolyte charge transfer). This approach was shown to yield quantitatively similar results as the conventional constant applied potential method, with the added benefit of employing simpler and faster 3-dimensional PBCs. The finite electric field {\bf E} method can also be extended to constant electric displacement Hamiltonians. The latter were introduced by Stengel, Spaldin and Vanderbilt to study ferroelectric systems using electronic structure calculations~\cite{stengel2009b}. Fixing the electric displacement corresponds to open-circuit electrical boundary conditions. Once applied to an electrochemical interface, it means that the potential is not anymore controlled, opening new perspectives for the simulation of electrochemical systems.

\begin{figure}[ht!]
	\begin{center}
 \includegraphics[width=0.8\textwidth]{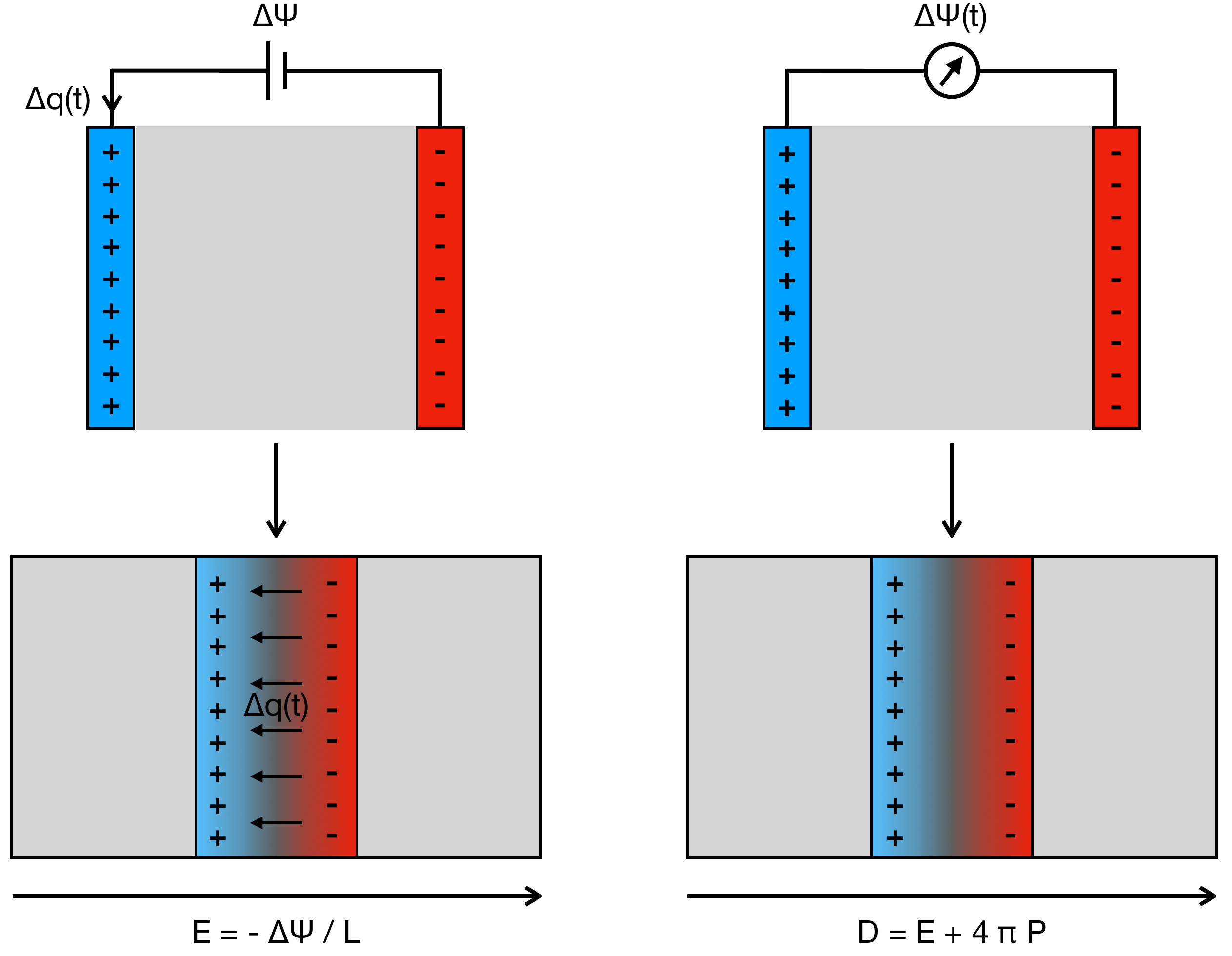}
\end{center}
	\caption{Top: electrochemical capacitors at fixed potential difference $\Delta \Psi$ (left) and under open circuit conditions (right). In the first case the charge is allowed to flow from one electrode to the other while on the second it is not. As a consequence the potential difference then fluctuates with time.  The light gray region contains the electrolyte. Bottom: Realization of these two setups using finite field method (left: the electric field {\bf E} is fixed; right: the electric displacement {\bf D} is fixed). Note that the difference in Poisson potential $\Delta \Psi$ can be directly identified with the difference in Galvani (inner) potential because, as imposed by the periodic boundary conditions, the electrolyte has the same composition on either end of the MD cell.}
\label{fig:schema}
\end{figure}

 In this work, we describe how constant electric displacement can be applied to a single metal slab to study nanoscale capacitors as schematized on Figure \ref{fig:schema}. As a first application of the method, we perform a computational amperometry study of a system made of a typical room-temperature ionic liquid in contact with electrified graphite. In analogy to electrochemical experiments, the electrical displacement is linearly increased in time, which leads to a build-up of electrode surface charge at a constant rate (see below). We study the voltage response, from which the capacitance of the system can be extracted and compared to conventional (constant applied potential) method. Varying the charging rate we can also obtain information about the relevant characteristic times.

The finite electric displacement field Hamiltonian is defined as~\cite{stengel2009b}
\begin{equation}
	H_D = H + \frac{\Omega}{8\pi}({\bf D}-4\pi{\bf P})^2
\label{eqn:hd}
\end{equation}	
\noindent where $H$ is the conventional Hamiltonian of the system including the kinetic and the potential energies, $\Omega$ is the volume of the supercell, ${\bf D}={\bf E}+4\pi{\bf P}$ is the electric displacement field, and {\bf P} the polarization per unit volume. It is worth noting that the latter depends on the choice of the periodic boundaries, an issue that is easily overcome by introducing the itinerant polarization for the electrolyte. The polarization ${\bf P}$ in Eq.~\ref{eqn:hd} is computed from the total dipole moment of the MD cell. Therefore, when a constant potential electrode is included ${\bf P}$ must also contain the contribution from the fluctuating electrode charges~\cite{dufils2019a}. In a system containing a conducting electrode, represented using fluctuating charges $\{q_i\}$ placed on the atomic sites, the potential $\Psi_i$ on each atomic site is given by
\begin{equation}
	\Psi_i = \frac{\partial U_c}{\partial q_i}-{\bf r}_i\cdot{\bf E}
\end{equation}	
\noindent where $U_c$ is the Coulombic contribution to the potential energy. The constant potential condition can be enforced within the electrode by using the conjugate gradient method as described in reference \citenum{dufils2019a}. The main difference here is that {\bf E} is not constant, hence neither is the Poisson potential across the cell $\Delta \Psi=-E L_z$ (where $L_z$ is the length of the box). This is consistent with the simulation of an electrochemical cell under open-circuit conditions. Note that we use here a hybrid method, where {\bf E} is not constant along the direction perpendicular to the electrode, while it is kept constant to {\bf 0} in the parallel direction, an approach similar to previous studies~\cite{zhang2016f,zhang2019a}.

\begin{figure}[ht!]
	\begin{center}
 \includegraphics[width=0.5\textwidth]{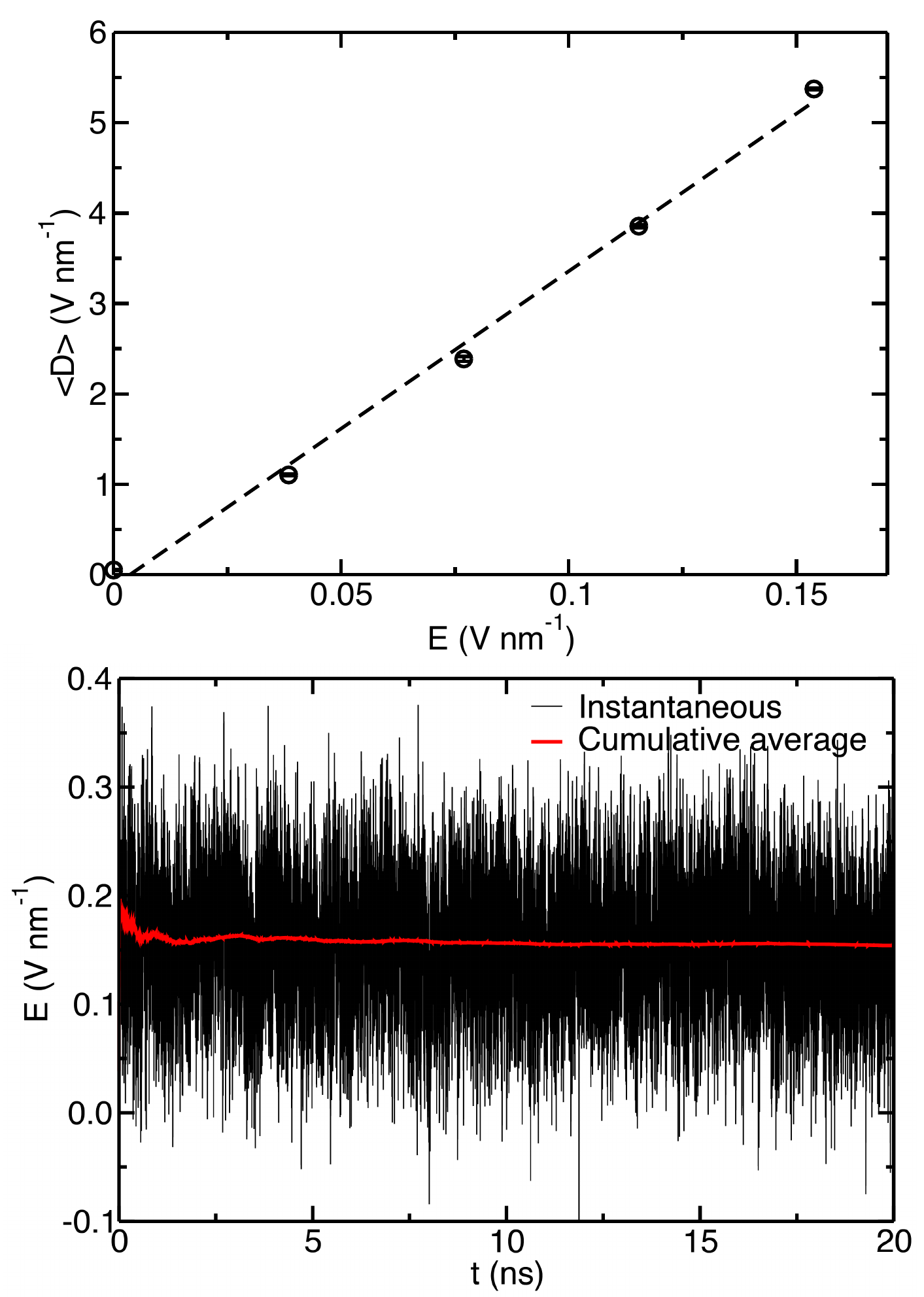}
\end{center}
	\caption{Top: average electric displacement calculated from the finite electric field simulations. The 95~\% confidence intervals are smaller than the symbol size, and a linear regression is shown as a dashed line. (Bottom: Cumulative average of the electric field along a simulation performed at finite electric displacement of 5.37~V~nm$^{-1}$ (which corresponds to the average value of $D$ during a simulation performed at fixed $E$ of 0.1538~V~nm$^{-1}$). The final value of $\langle E\rangle$ is of 0.1540~V~nm$^{-1}$. The confidence interval is not shown for clarity reasons but the standard deviation of 0.00045~V~nm$^{-1}$ is obtained for the final value of the cumulative average.}
\label{fig:field}
\end{figure}

In order to check the feasibility of the method, we have simulated a nanocapacitor made of a single graphite electrode and the 1-ethyl-3-methylimidazolium hexafluorophosphate ionic liquid for the electrolyte. This system was chosen because its electrochemical characteristics were already studied using constant applied potential MD simulations.~\cite{merlet2011a} We have performed a first series of calculations at finite {\bf E}, from which we have determined the average electric displacement $\langle D \rangle (E)$. As shown in Figure~\ref{fig:field}, the variation is almost linear. Then the corresponding values of $D$ were fixed in a second series of simulation. Figure \ref{fig:field} displays the cumulative average of $E$ along the simulation for the largest applied value. The final value of the average field $\langle E \rangle$ is of 0.1540~V~nm$^{-1}$, which matches well with the initially fixed value of 0.1538~V~nm$^{-1}$. The integral capacitances obtained at finite {\bf E} and {\bf D} also yield similar values.  

\begin{figure}[ht!]
	\begin{center}
 \includegraphics[width=0.8\textwidth]{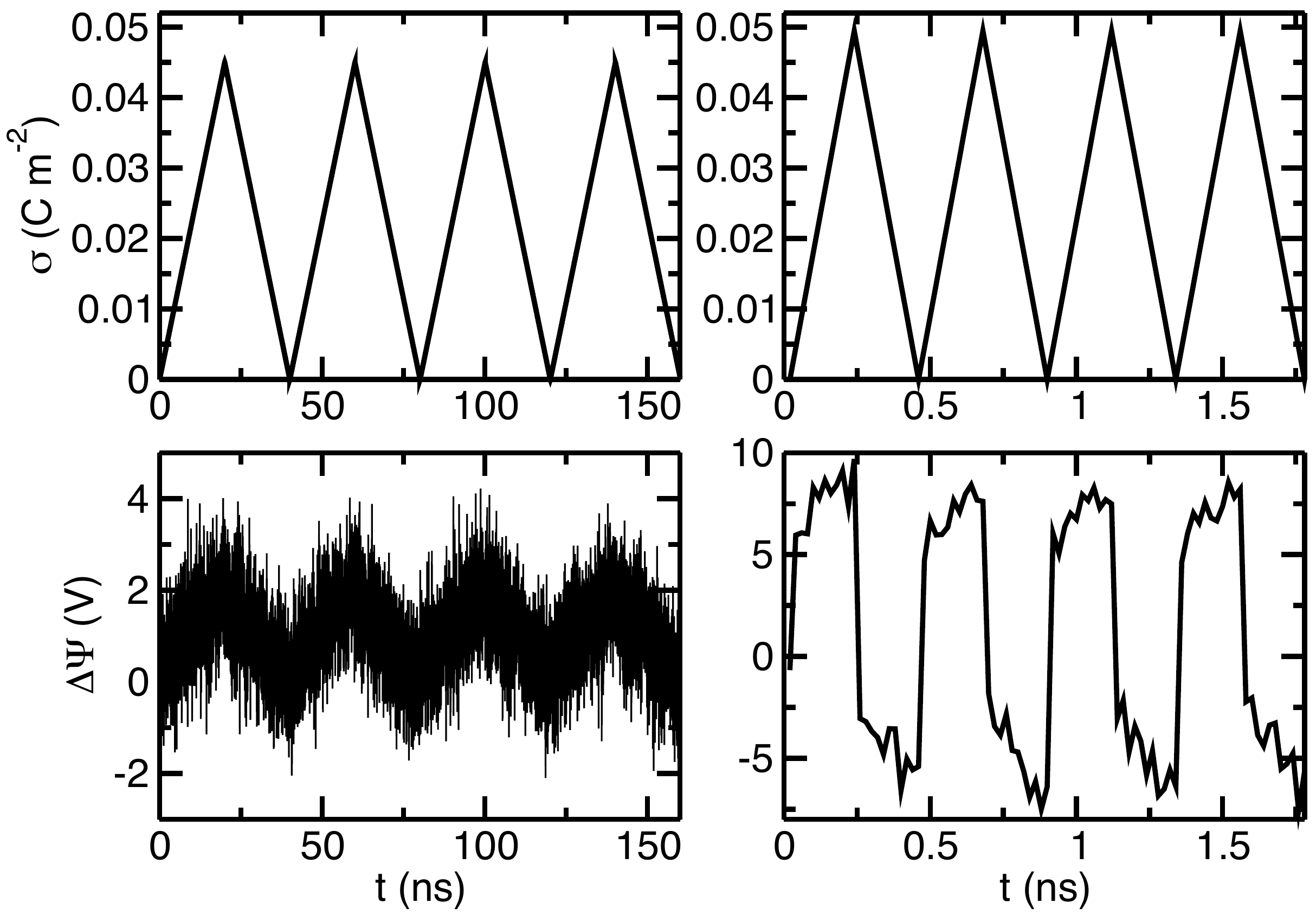}
\end{center}
	\caption{Computational amperometry experiment: Variation of the surface charge (top) and of the potential (bottom) for a slow (0.0077~V~nm~${-1}$~s$^{-1}$, left) and a fast (0.77~V~nm~${-1}$~s$^{-1}$, right) electric displacement ramps. }
\label{fig:amperometry}
\end{figure}

The surface charge $\sigma$ of the electrode is controlled by the value of the electric displacement according to $D=4\pi \sigma$  as verified by the test calculation reported in the Supplementary Information. The explanation for this simple relation is that in our constant dielectric displacement scheme all charge, including the ionic mobile charge  in the electrolyte, is accounted for by the polarization ${\bf P}$.  The dielectric displacement $D$ in the planar slab geometry is therefore the same everywhere in the cell.  In the metal electrode $E=0$ and hence $D=4 \pi P = 4 \pi \sigma$. Thus performing several consecutive simulations with a linear increase (decrease) of {\rm D} allows to study the response of the system to the application of a constant current. In electrochemistry experiments, this technique called amperometry is routinely used to characterize supercapacitors. The potential $\Delta \Psi$ is expected to vary with time $t$ according to
\begin{equation}
	\Delta \Psi = i\left(R+\frac{t}{C}\right) \label{eq:UI}
\end{equation}
\noindent where $i$ is the current, $R$ is the so-called equivalent series resistance (ESR) and $C$ the capacitance of the device.~\cite{portet2005a} The ESR contains contributions form the current collector, the electrode, the separator, the electrolyte and all the interfaces between them. In our case, we only expect terms arising from the electrolyte and its interface with the electrode since the latter is modelled as a conductor and the current collector and separator are not included in the simulations. Both $R$ and $C$ vary with the applied current because the system is not under equilibrium conditions. 

Here we performed four series of simulations with varying electric displacement ramps (0.269, 2.69, 5.38 and 26.9~V~nm$^{-1}$~ns$^{-1}$, which were calculated so that the potential should reach approximately 2~V in 20, 4, 2 and 0.2~ns respectively under equilibrium conditions). The two extreme cases are displayed on Figure \ref{fig:amperometry}, where the surfacic charge $\sigma$ of the positively charged surface of the slab and the voltage are both plotted with respect to the simulation time for four successive charge/discharge cycles. At low current (slow {\bf D} ramp), the potential, albeit experiencing large fluctuations, increases and decreases linearly with the surface charge, which corresponds to a system responding to the perturbation in a quasi-equilibrium state. At high current, a large overpolarization is observed as soon as the charge starts to increase or decrease, which is due to the ESR. It shows that the system is not able to respond fast enough to the perturbation through the diffusion of cations and anions. After this overpolarization, the potential increases/decreases importantly (compared to the low current case) with time, which points towards a lower capacitance according to equation \ref{eq:UI}. 

\begin{figure}[ht!]
	\begin{center}
 \includegraphics[width=0.8\textwidth]{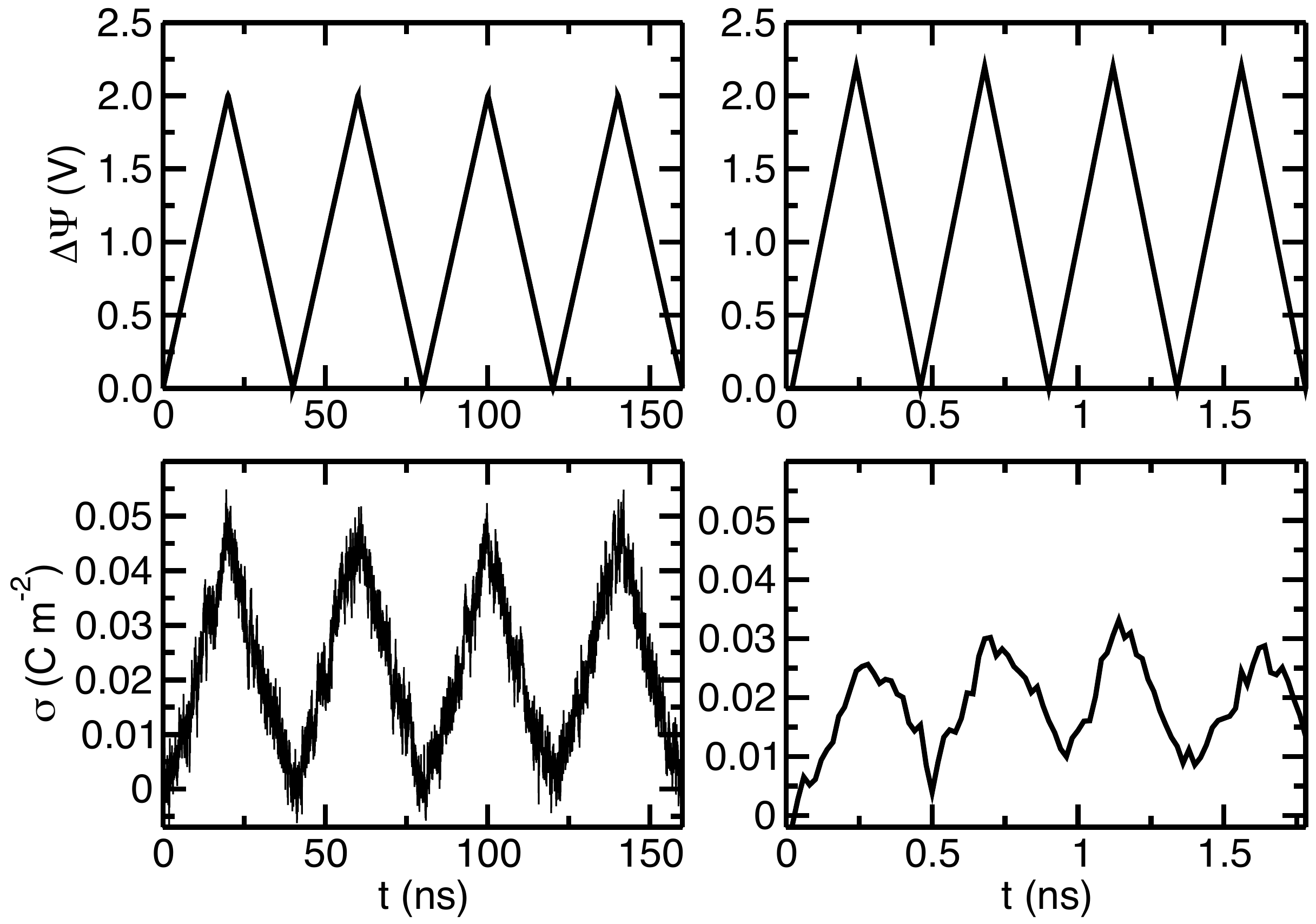}
\end{center}
	\caption{Computational voltammetry experiment: Variation of the the potential (top) and of the surface charge (bottom)  for a slow (0.0077~V~nm~${-1}$~s$^{-1}$, left) and a fast (0.77~V~nm~${-1}$~s$^{-1}$, right) electric displacement ramps.}
\label{fig:voltammetry}
\end{figure}

Such a behavior was already observed by He {\it et al.} in a computational voltammetry study, in which the applied voltage was successively increased/decreased at various rate.~\cite{he2015a} In order to compare the response of nanoscale capacitors to voltage/charge ramps, we have performed a similar study where the electric field was progressively increased at four different rates (0.0077, 0.077, 0.154 and 0.77~V~nm$^{-1}$~ns$^{-1}$) corresponding to the same conditions as in the amperometry calculations described above. As shown on Figure \ref{fig:voltammetry}, at low rate the response is linear as in the amperometry case, and at high rate the non-equilibrium conditions result in a lower charge accumulated at the surface (hence also to a lower capacitance). 

\begin{figure}[ht!]
	\begin{center}
 \includegraphics[width=0.8\textwidth]{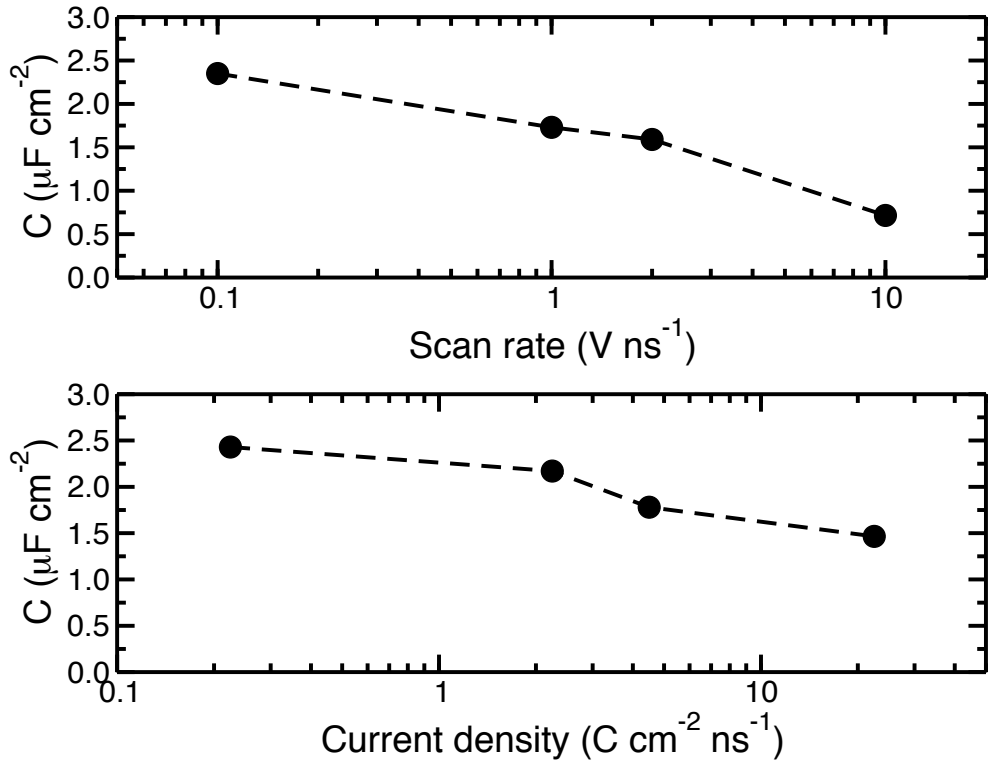}
\end{center}
	\caption{Capacitances extracted from the computational voltammetry (top) and amperometry (bottom) experiments, with respect to the voltage scan rate and of the applied current density, respectively. }
\label{fig:capacitance}
\end{figure}

On Figure \ref{fig:capacitance} we compare the capacitances obtained in the voltammetry and amperometry simulations. The rates and intensities correspond to similar simulation times so the behavior of the two plots can be compared (note that these quantities were used instead of the finite field ramp values in order to have consistency with the corresponding experiments). In both cases, the equilibrium capacitance (2.37~$\mu$F~cm$^{-2}$) is well recovered at low scan rate or current. The decrease of the capacitance is then much larger  in the voltammetry than in the amperometry. This is consistent with previous results obtained in constant {\bf E} and {\bf D} simulations performed on charged insulator/electrolyte interfaces.~\cite{zhang2016f} This is due to the fact that the polarization fluctuations are more than one order of magnitude shorter in the constant {\bf D} ensemble. Here this one order of magnitude difference is recovered since the capacitance obtained at the largest current (1.5~$\mu$F~cm$^{-2}$) matches well with the one obtained for simulations at constant {\bf E} with 5 to 10 times longer simulation times (scan rates of 1 and 2~V~ns$^{-1}$).

In conclusion, we have shown in this work how the finite electric displacement method can be combined with a constant potential electrode in order to simulate electrochemical systems under open circuit conditions.  We have then shown that it is possible to perform computational amperometry experiments by applying electric displacement ramps for charging and discharging a nanoscale capacitors. At slow rate, the system is able to respond while staying under equilibrium conditions and the full capacitance is reached, but at high rate the ions do not diffuse rapidly enough, leading to a large overpolarization similar to the equivalent series resistance observed in experiments. Nevertheless, the obtained capacitance remains higher than in the case of computational voltammetry experiments under similar conditions. Other application of the method could be the simulation of systems with a fixed total charge on the electrode, but still allowing charge reorganization in response to the electrolyte fluctuations. In addition, simulations of nanoporous electrode-based capacitors are notoriously slow to equilibrate~\cite{breitsprecher2018a,breitsprecher2020a}, a situation that could be improved by adopting the present finite field calcualations to the case of complex electrode geometries.

\section*{Methods}

The nanoscale capacitor is composed of a graphite electrode made of 2912 carbon atoms (which are kept fixed, forming 7 graphitic planes) and 320 ion pairs of 1-ethyl-3-methylimidazolium hexafluorophosphate. The system is simulated  in the canonical ensemble at a temperature of 400~K,  using a Nos\'e-Hoover thermostat chain~\cite{martyna1992a} with a relaxation time of 125~fs. A timestep of 2~fs is used to integrate the equations of motions, and the carbon atom charges are calculated at each timestep using a conjugate gradient procedure. The coarse-grained model developed by Roy and Maroncelli is used for the ionic liquid,~\cite{roy2010a} and the Lennard-Jones parameter for carbon was taken from Ref.~\citenum{cole1983a}. This combination of parameters was thoroughly tested for the study of such capacitors.~\cite{merlet2011a} The initial configuration was extracted from a simulation equilibrated using a two-electrodes setup, in which the two electrodes were merged and an additional graphitic plane was added. The lengths of the simulation cell are 32.24~\AA\, 34.37~\AA\ and 130.00~\AA\ along the $x$, $y$ and $z$ directions respectively. Two series of simulations were performed: A first one in which the electric displacement was systematically varied along a ramp. Different rates of 0.269, 2.69, 5.38 and 26.9~V~nm$^{-1}$~ns$^{-1}$ were used, in which the field was incremented/decremented every 20~ps, with corresponding total simulation times of 20, 4, 2 and 0.2~ns. In the second series of simulations, the electric field was progressively increased at four different rates (0.0077, 0.077, 0.154 and 0.77~V~nm$^{-1}$~ns$^{-1}$), with similar total simulation times as for the electric displacement simulations. Four successive charge/discharge cycles were performed in each case. All the simulations were performed with the MetalWalls simulation software.~\cite{marinlafleche2020a} 

\section*{Acknowledgements}
\noindent This project has received funding from the European Research Council  under the European Union's Horizon 2020 research and innovation programme (grant agreement No. 771294). This work was supported by the French National Research Agency (Labex STORE-EX, Grant  ANR-10-LABX-0076 and project SELFIE, Grant ANR-17-ERC2-0028). The authors acknowledge HPC resources granted by GENCI (resources of CINES, Grant No A0080910463).

\providecommand{\latin}[1]{#1}
\makeatletter
\providecommand{\doi}
  {\begingroup\let\do\@makeother\dospecials
  \catcode`\{=1 \catcode`\}=2 \doi@aux}
\providecommand{\doi@aux}[1]{\endgroup\texttt{#1}}
\makeatother
\providecommand*\mcitethebibliography{\thebibliography}
\csname @ifundefined\endcsname{endmcitethebibliography}
  {\let\endmcitethebibliography\endthebibliography}{}

\newpage

\section*{Supplementary Information}
In previous studies, it has been shown that for insulator-electrolyte interface, the net surface charge $\sigma$ is given by
\begin{equation}
D=4\pi\sigma \label{Dvssigma}
\end{equation}

\subsection{Simulations at constant electric displacement}
\begin{figure}[h!]
\begin{center}
\includegraphics[scale=1.0]{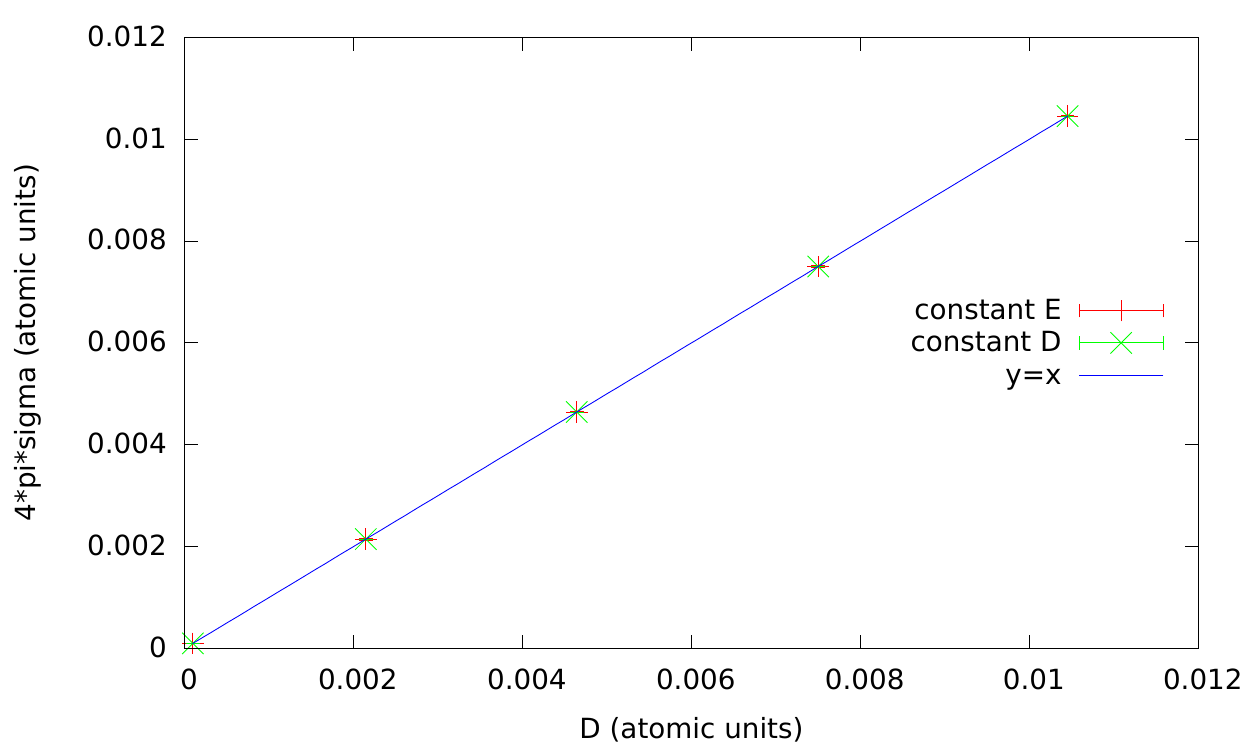}
\end{center}
\caption{Electrode surface charge in the finite D ensemble at thermodynamic equilibrium \label{surface_charge}}
\end{figure}

We first tested the extension of this relation to a metallic electrode in equilibrium molecular dynamics simulations. After 2 ns of equilibration, the average surface charge of the electrode is computed over 20 ns. The results are displayed on figure \ref{surface_charge} and show that this relation is verified in the applied voltage range.

\subsection{Electric displacement ramps}
\begin{figure}

\includegraphics[scale=1.2]{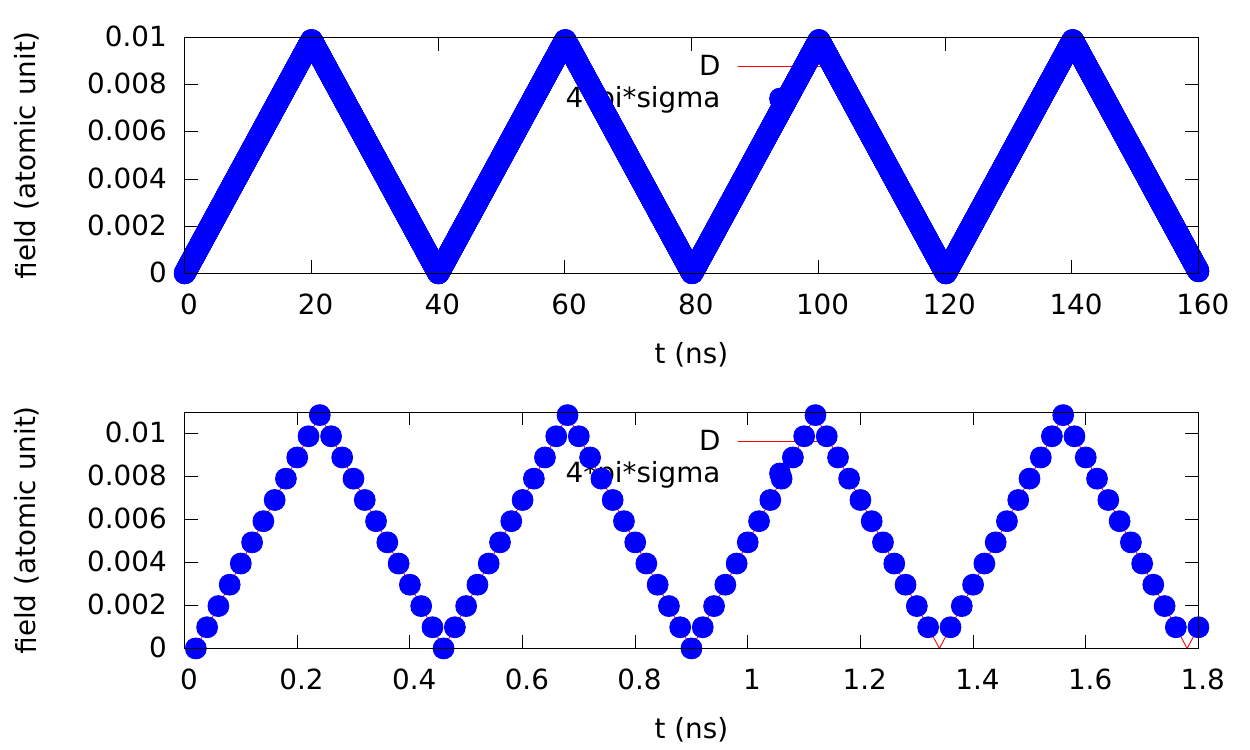}
\caption{Evolution in time of the electric displacement and the electrode surface charge for electric displacement ramps for 0.1V/nm (top) and 10.0 V/nm (bottom) \label{DvsSigma}}

\end{figure}

During the electric displacement ramps, we recorded the evolution of the electrode surface charge to check the value of the current. For this, we represent the evolution in time of $4\pi\sigma$, along with D on figure \ref{DvsSigma} for the two extreme values of the applied current (corresponding to a scan rate for the applied voltage of 0.1 V/ns and 10 V/ns). From this graph we conclude that relation \ref{Dvssigma} applies even in non equilibrium situations. This is also the case for the other current values we used. This confirms that these ramps correspond to simulations at a constant current. Note that the surface charge is averaged over 20 ps. This doesn't mean however that this relation is verified for every single configuration.

\end{document}